\documentclass[
  aps, pra, reprint, twocolumn, superscriptaddress, floatfix, longbibliography
]{revtex4-2}

\usepackage{amsmath,amssymb,bm}
\usepackage{graphicx}
\usepackage{xcolor}
\usepackage[colorlinks=true,allcolors=blue]{hyperref}

\newcommand{\HeadlineRatio}{1.440}
\newcommand{\HeadlineSpread}{0.011}
\newcommand{\HeadlineE}{0.905}
\newcommand{\HeadlineH}{1.053}
\newcommand{\HeadlineR}{0.860}
\newcommand{\HeadlineP}{0.0699}
\newcommand{\HeadlineDiag}{0.425}
\newcommand{\HeadlineCoh}{0.479}
\newcommand{\HeadlineCohPercent}{53}

\begin{document}

\title{Collective advantage from a minimal record in a quantum information engine}

\author{Kangqiao Liu}
\email{kqliu@xhu.edu.cn}
\affiliation{School of Science, Xihua University, Chengdu 610039, China}
\affiliation{Key Laboratory of High Performance Scientific Computation,
             Xihua University, Chengdu 610039, China}

\author{Jie Gu}
\email{jiegu1989@gmail.com}
\affiliation{Chengdu Academy of Educational Sciences, Chengdu 610036, China}

\author{Deyou Chen}
\email{deyouchen@hotmail.com}
\affiliation{School of Science, Xihua University, Chengdu 610039, China}
\affiliation{Key Laboratory of High Performance Scientific Computation,
             Xihua University, Chengdu 610039, China}

\date{\today}

\begin{abstract}
For a single particle, a quantum information engine can turn measurement fluctuations into
transport by raising a barrier behind the particle each time its position is measured. When
many particles are present, the barrier still depends on just one number, the position of the
leftmost particle, so we let the demon measure that order statistic and nothing else. A demon
that resolves every position instead pays a record entropy that grows with particle number, even though the extra distinctions never change where the barrier is placed. We show
that the coarser measurement supplies an energy that is bounded independently of particle
number, and that this bound yields a ceiling on the record entropy that falls as the filling
increases. At the same tilt, the collective engine then delivers $44\%$ more work per recorded
nat than independent single-particle engines, each with its own optimized cycle time and each
running at equal or greater power per particle. Pauli blocking ends this advantage once the
accessible region just above the wall fills. We also evaluate the coherence that the coarse measurement leaves behind, and
the cost of placing the barrier imprecisely.
\end{abstract}

\maketitle

% ============================================================================
\section{Introduction}
\label{sec:intro}

Information engines convert measurement and feedback into useful work, and the value of that
feedback is limited by the information the demon must eventually erase
\cite{Szilard1929,Brillouin1951,Landauer1961,Bennett1982,Maruyama2009,Parrondo2015}. Feedback
enters the second law through the information that the controller acquires
\cite{Sagawa2008,Sagawa2009,SagawaUeda2010,Jacobs2009,Faist2015,WisemanMilburn2009,Seifert2012},
and the erasure cost itself has by now been measured directly in single-particle experiments
\cite{Berut2012,Jun2014} and sharpened for reservoirs of finite size
\cite{Esposito2011,ReebWolf2014,Goold2015}. When the demon writes to an explicit tape, this
accounting becomes a statement about the entropy of the recorded sequence
\cite{MandalJarzynski2012,BaratoSeifert2013,Boyd2016,Horowitz2014,Strasberg2017}, which is the
form we use throughout.

Quantum information engines have been proposed and realized on several platforms
\cite{Kim2011,Park2013,Brandner2015,Elouard2018,Toyabe2010,Koski2014,Koski2015,Chida2017,
Paneru2018,Ribezzi2019}, among them photonic \cite{Vidrighin2016}, superconducting
\cite{Cottet2017,Masuyama2018,Naghiloo2018}, and nuclear-spin \cite{Camati2016} devices. In the
quantum case the measurement is more than a source of information, since it can itself supply
the energy that the cycle goes on to extract
\cite{Elouard2017,Manzano2018,Funo2013,Gong2016}.

One such engine is purely spatial. Reference~\cite{Liu2026} places a single particle on a
tilted one-dimensional lattice, measures its position, and raises a barrier immediately behind
it. Repeating that cycle rectifies quantum measurement fluctuations into transport against the
tilt. The contrast with the undriven problem is sharp, because without the demon the particle
merely performs Bloch oscillations and travels nowhere over long times
\cite{Bloch1929,Wannier1960,Gluck2002,Hartmann2004}. Cold atoms realize those oscillations
cleanly \cite{BenDahan1996,Wilkinson1996,Anderson1998,Morsch2001,Ferrari2006}, and they persist
even without a lattice \cite{Meinert2017}. Site-resolved experiments have shown, moreover, that
few-particle quantum walks in tilted lattices develop correlations and collective Bloch
dynamics well beyond the single-particle limit \cite{Preiss2015}. Extending the transport demon
past one particle is therefore a natural place to ask how much of a many-body state the
feedback actually has to resolve.

For many particles, the most direct extension is to measure every position, identify the
leftmost particle, and apply the same feedback as before. That extension creates a mismatch at
once. A full configuration distinguishes a great many outcomes, yet every configuration sharing
a leftmost position calls for the same barrier, so the record grows expensive while the added
distinctions change nothing about the action taken. Keeping only what the controller acts on
has a long history in information thermodynamics and in partially observable engines
\cite{Still2012,Bergli2014,StillDaimer2022,Still2020}. In stochastic thermodynamics this is the
role played by a sufficient statistic \cite{Matsumoto2018}, and the information passing between
the measured system and its controller obeys a balance of its own
\cite{Ito2013,Hartich2014,HorowitzSandberg2014}. Quantum Szilard engines and many-particle
demons show, in addition, how particle statistics reshapes information-to-work conversion
\cite{Kim2011,Hlousek2024}. What repeated spatial transport adds is a single dynamical model in
which the information written to the tape, the measurement backaction, and the collective
motion can all be compared.

In the present engine the controlling variable is unusually simple. Wherever the particles
happen to sit, the barrier goes immediately below the leftmost one, so the feedback depends on
the entire configuration only through that single position. Being the smallest of the occupied
site labels, it is an order statistic \cite{DavidNagaraja2003}. We therefore compare two
measurements that produce the very same barrier action. The full-configuration measurement resolves the
complete configuration, whereas the order-statistic measurement resolves the leftmost position
alone and leaves configurations sharing it unresolved. The two consequently differ both in what
they record and in the state they leave behind for the next cycle. That difference cannot be
reproduced by measuring everything and discarding the rest afterwards, because by then the
finer measurement has already changed the quantum state.

We find that the energy an order-statistic measurement supplies is bounded independently of
the number of particles, and that this bound puts a ceiling on the entropy of the record that
falls as the lattice fills. At a calibrated $^6$Li operating point the engine then delivers
$44\%$ more work per nat of recorded information than the same number of independent
single-particle engines at the same tilt. Each of those runs at the same or greater power per
particle. The full-configuration demon, by contrast, stays below that reference at both tilts
we examined and for two to eight particles, and about $95\%$ of its ten-particle record resolves distinctions
that never change the barrier. The gain is non-monotonic in particle number, since Pauli
blocking suppresses the wall once the accessible region just above it fills. A controlled dephasing comparison
shows, finally, that the coherence surviving the coarse measurement adds to the work without
being necessary for the advantage.

In Sec.~\ref{sec:setup} we define the cycle, the two demons, and the single-particle
reference. Section~\ref{sec:bounds} derives the bound on the measurement energy and the
entropy ceiling that follows from it. Section~\ref{sec:advantage} compares the record cost of
the two demons and reports the collective advantage. Section~\ref{sec:scaling} shows how Pauli
blocking limits that advantage. Section~\ref{sec:coherence} separates the saving in record cost
from the coherence the coarse measurement retains. Section~\ref{sec:errors} examines imprecise
barrier placement. Section~\ref{sec:experiment} gives a $^6$Li lattice realization.

\section{Setup}
\label{sec:setup}

\begin{figure}[tbp]
  \centering
  \includegraphics[width=\columnwidth]{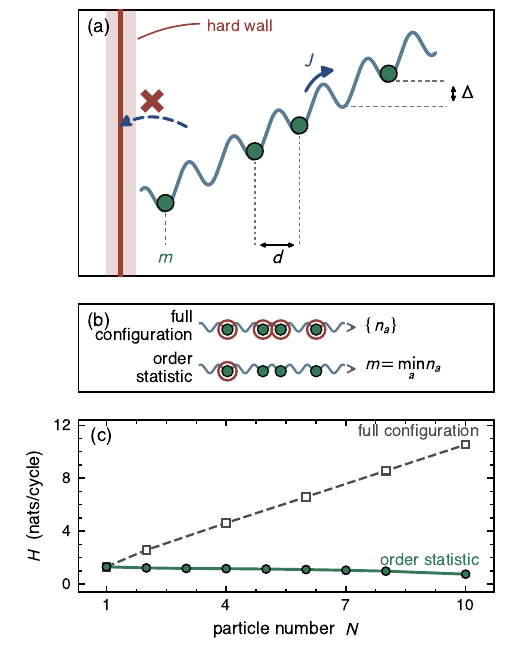}
  \caption{%
    The full-configuration measurement resolves distinctions that never change the barrier action, while
    the order-statistic measurement records the variable that controls it. (a)~$N$ fermions on
    a tilted lattice of spacing $d$ and step height $\Delta$, hopping with amplitude $J$. After
    each measurement a hard wall is placed below the leftmost particle at $m$, so that hops
    across it are forbidden outright rather than merely suppressed. (b)~The record
    each demon takes from the same configuration. The full-configuration demon resolves every
    occupied site, while the order-statistic demon resolves only the leftmost. (c)~Record entropy rate at
    $\alpha=0.15$ and $\tau=1.85$. The full-configuration record grows nearly linearly with particle
    number. The order-statistic record stays near one nat and decreases as the minimum
    concentrates. Each curve is evaluated on the histories generated by its own measurement.}
  \label{fig:setup}
\end{figure}

Consider $N$ spin-polarized fermions on a one-dimensional tilted lattice of $M$ sites, closed
on the left by a rigid wall, as shown in Fig.~\ref{fig:setup}(a). Everything below is written
in the frame that moves with the wall, so that the lowest site the particles may occupy is
always labelled $j=0$, however far the wall has travelled. We call a state written in this
moving frame a \emph{wall-frame state}. The Hamiltonian on the half line is
\begin{equation}
 \begin{aligned}
 \hat H&=\hat K+\Delta\hat X,\qquad
 \hat X=\sum_{j\ge0}j\hat n_j,\\
 \hat K&=-J\sum_{j\ge0}(\hat c^\dagger_{j+1}\hat c_j+\mathrm{H.c.}).
 \end{aligned}
 \label{eq:hamiltonian}
\end{equation}
where $J$ is the nearest-neighbor hopping amplitude and $\Delta$ is the energy step between
neighboring sites, while $\hat c_j$ and $\hat c^\dagger_j$ annihilate and create a fermion at
site $j$ and $\hat n_j\equiv\hat c^\dagger_j\hat c_j$ counts its occupation. The first term
$\hat K$ moves particles along the chain. The second is their potential energy, written through
the total position $\hat X\equiv\sum_j j\hat n_j$, so that the product $\Delta\hat X$ is an
energy and not a change in $\hat X$. Throughout the paper we measure energies in units of $J$,
powers in units of $J^2/\hbar$, and time through the dimensionless combination
$\tau\equiv Jt/\hbar$. A single dimensionless parameter is then left, the tilt
$\alpha\equiv\Delta/J>0$, and the numerical results below use Eq.~\eqref{eq:hamiltonian} with
no interaction between the fermions.

To convert measurement into transport, the engine repeats a four-step cycle.

\emph{Step 1, measurement.} The demon measures the particle positions, either in full or only
through the position of the leftmost particle.

\emph{Step 2, feedback.} A hard wall is placed on the site immediately below the leftmost
particle, so that no particle can hop downward across the new boundary.

\emph{Step 3, unitary evolution.} The state evolves under Eq.~\eqref{eq:hamiltonian} for a
time $\tau$.

\emph{Step 4, recording.} The measurement outcome is appended to the demon's tape, which is
kept during the run and erased once the run is over.

Only Step 2 consumes the outcome, and it consumes very little of it. Writing a configuration as
the ordered list $\boldsymbol n=(n_1<\cdots<n_N)$ of occupied sites, the wall is placed below
\begin{equation}
 m=n_1=\min_a n_a,
 \label{eq:sufficient}
\end{equation}
so that two configurations sharing the same $m$ receive the same feedback, whereas two with
different $m$ call for the wall in different places. This is the gap the paper exploits, and it
separates the two demons we compare. Both follow the same four steps and differ only in Step 1.
The \emph{full-configuration (FC) demon} resolves the entire configuration and records
$\boldsymbol n$, which leaves the particles in a Fock state. The \emph{order-statistic (OS)
demon} resolves only $m$, and therefore leaves untouched whatever superposition the cycle
has built up among configurations that share that minimum. Figure~\ref{fig:setup}(b) contrasts the two records.

Consider now one cycle of the OS demon. A cycle begins immediately after feedback,
when the wall-frame state $\rho_t$ has the origin occupied with certainty,
\begin{equation}
 \hat n_0\rho_t=\rho_t.
 \label{eq:occupied_origin}
\end{equation}
Free evolution for a time $\tau$ carries this state to
$\sigma_t=\mathcal U_\tau\rho_t\mathcal U_\tau^\dagger$, where $\mathcal U_\tau$ is the unitary
generated by Eq.~\eqref{eq:hamiltonian}. The demon then locates the leftmost occupied site with
the projectors
\begin{equation}
 P_m=\hat n_m\prod_{j<m}(1-\hat n_j),\qquad
 p_t(m)=\mathrm{Tr}(P_m\sigma_t),
 \label{eq:minimum}
\end{equation}
which report the outcome $m$ with probability $p_t(m)$. Each $P_m$ projects onto a whole family
of configurations rather than onto a single one, so the outcome alone does not fix the state
that the measurement leaves behind. We take the measurement to be of the L\"uders type \cite{Luders1950,WisemanMilburn2009}, which
is the standard choice that disturbs the state no more than the outcome demands. The conditioned state
is then $P_m\sigma_tP_m/p_t(m)$, and every coherence internal to the sector survives. The wall is then advanced by $m$ sites and the coordinates are translated by $T_m$,
which returns the first allowed site to the origin. On the half line $T_m$ acts as
$T_m|n_1,\ldots,n_N\rangle\equiv|n_1-m,\ldots,n_N-m\rangle$ for configurations with
$n_1\ge m$, so it is an isometry on the range of $P_m$ rather than a translation of the
whole space, and $T_m^\dagger$ restores the laboratory coordinates. One conditioned cycle therefore reads
\begin{equation}
 \rho_{t+1}=
 \frac{T_mP_m\mathcal U_\tau\rho_t\mathcal U_\tau^\dagger P_mT_m^\dagger}{p_t(m)}.
 \label{eq:cycle}
\end{equation}
Because the wall is inserted into a region that is empty by construction, its ideal on-site
potential does no mean work while it is relocated, and the translation $T_m$ only resets the
frame. In the laboratory frame, every particle keeps the displacement it has accumulated from
the moving wall.

The two demons differ in what they must write down. At a fixed conditioned premeasurement
state, any exact record $Y_t$ from which the demon can reconstruct the barrier position
satisfies
\begin{equation}
 H(Y_t\mid Y_{<t})\ge H(m_t\mid Y_{<t}),
 \label{eq:information}
\end{equation}
where $m_t$ is the leftmost position found in cycle $t$, $H(\cdot\mid\cdot)$ is the conditional
Shannon entropy, and $Y_{<t}$ denotes everything recorded in the cycles before $t$. Conditioning on that past is essential, because the earlier records
fix the state on which the present measurement acts. Recording $m_t$ alone attains that lower bound, and is
therefore the minimal record the feedback admits. Any finer record sits above it by whatever
the full configuration says beyond its minimum. The two measurements also leave different states behind, so their stationary
entropy rates have to be evaluated on their own histories rather than subtracted as though one
process produced both.

Because the Hamiltonian is quadratic, both measurements carry a Slater determinant to a Slater
determinant, so each demon can be propagated exactly on its own conditioned history. It is
enough to follow the $N$ occupied one-particle orbitals, collected as the columns of an
$M\times N$ matrix $\Phi$, from which the one-body density matrix
$\Gamma=\Phi\Phi^\dagger$ carries every quantity we report below.
Appendix~\ref{app:gaussian} gives that reduction and its many-body checks.

For a run of $L$ cycles, the chain rule gives
\begin{equation}
 H(Y_{1:L})=\sum_{t=1}^{L}H(Y_t\mid Y_{<t}).
 \label{eq:entropy_rate}
\end{equation}
We erase the tape after the run, so the stationary information cost per cycle is the entropy
rate
$H=\langle-\ln p(Y_t\mid Y_{<t})\rangle$ in nats, the mean of the surprisal $-\ln p$
carried by one recorded outcome. We denote the OS and FC
rates by $H_{\min}$ and $H_{\mathrm{full}}$.

What the engine produces is most easily seen in the laboratory frame, where the wall has
actually moved. Over $L$ cycles it advances by $S_L\equiv\sum_{t=1}^L m_t$ sites in all,
carrying the particles up the tilt with it, and the potential energy they gain is
\begin{equation}
 W_L=\Delta\left[N S_L+
    \langle\hat X\rangle_{\rho_{L+1}}-
    \langle\hat X\rangle_{\rho_1}\right].
 \label{eq:finite_work}
\end{equation}
The first term counts the energy stored by moving the wall, since each of the $N$ particles
rises by $\Delta$ for every site the wall advances. The remaining two terms correct for the
cloud rearranging itself relative to the wall, which matters over a finite run but averages
away in a stationary state. Dividing by $L$ therefore leaves the stored work per cycle,
\begin{equation}
 E=\Delta N\langle m\rangle.
 \label{eq:wall_work}
\end{equation}

That energy comes from the measurement. Unitary evolution conserves the energy of
Eq.~\eqref{eq:hamiltonian}, and the wall does no mean work because it is placed where there is
nothing to push. Averaging over outcomes while keeping no record turns a state $\sigma$ into
$\sum_mP_m\sigma P_m$, and the energy this costs,
\begin{equation}
 Q_{\mathrm{meas}}\equiv\mathrm{Tr}\Big\{\hat H\Big[\sum_mP_m\sigma P_m-\sigma\Big]\Big\},
 \label{eq:heat}
\end{equation}
is what the demon injects each time it looks. Averaging the feedback translation over outcomes
gives
\begin{equation}
 Q_{\mathrm{meas}}=\Delta N\langle m\rangle+
 \langle\hat H\rangle_{\rho_{t+1}}-
 \langle\hat H\rangle_{\rho_t}.
 \label{eq:balance}
\end{equation}
In a stationary cycle the last two terms cancel, and the balance closes on itself,
\begin{equation}
 E=\Delta N\langle m\rangle=Q_{\mathrm{meas}}.
 \label{eq:stationary}
\end{equation}
Every unit of energy the engine stores is therefore one the measurement supplied, which is what
makes the bound of Sec.~\ref{sec:bounds} so restrictive. Appendix~\ref{app:energy} gives the
derivation, and the conditions under which position-diagonal interactions leave it intact.

Running the engine costs what the record will eventually cost to erase, so the ideal
efficiency of a cycle is the stored work divided by that work plus the erasure cost of the
record \cite{Liu2026}. At a fixed erasure temperature, this efficiency rises and falls with a
single ratio, the work the engine stores per nat it writes down,
\begin{equation}
 R\equiv\frac{E}{H},
 \label{eq:R}
\end{equation}
at fixed erasure temperature. We report $R$ together with the power per particle
$\mathcal P=JE/(N\hbar\tau)$, written in script to keep it clear of the outcome
probabilities $p_t(m)$. A concrete realization also has control and readout costs. Those enter in
Sec.~\ref{sec:experiment}, while $R$ isolates the conversion of measurement energy and recorded
information in the ideal cycle.

We compare the collective engine with $N$ independent copies of the single-particle engine at
the same tilt. Because a single-particle engine can trade power for work per nat by changing its
cycle time, the reference is the envelope
\begin{equation}
 R_{\mathrm{ref}}(\mathcal P)=\sup_{\tau':\,\mathcal P_1(\tau')\ge\mathcal P}R_1(\tau').
 \label{eq:reference}
\end{equation}
A ratio $R_N/R_{\mathrm{ref}}>1$ means that the collective engine produces more work per recorded
nat than every single-particle operating point with equal or greater power per particle. For an
arbitrary single-particle cycle time,
$R_1(\tau)/R_{\mathrm{ref}}[\mathcal P_1(\tau)]\le1$, with equality on the optimized envelope.
Appendix~\ref{app:reference} constructs the envelope and verifies that the time windows used
below contain all power-feasible single-particle points.

\section{Exact relations for the cycle}
\label{sec:bounds}

The OS measurement does more than cut down the number of outcomes the demon has
to record. Because the minimum is a boundary coordinate, a nearest-neighbor hop can change it only when
the leftmost ordered particle itself moves. That single observation constrains both the energy
the measurement supplies and the entropy of the record it produces. In the same
spirit, the L\"uders measurement removes only those hopping coherences that connect different
values of the minimum, instead of dephasing the cloud throughout its interior. We derive the
two constraints in turn.

\subsection{Bound on the measurement energy}

We first bound the energy the measurement can inject in a single cycle. That bound does not
grow with the number of particles. Every term of
Eq.~\eqref{eq:hamiltonian} that is diagonal in position commutes with all the $P_m$, so the energy change in Eq.~\eqref{eq:heat} comes entirely from hopping. What the measurement
actually removes is the part of $\hat K$ that links different minimum sectors, namely
$\hat K-\sum_mP_m\hat KP_m$, and $Q_{\mathrm{meas}}$ is minus its expectation value. In the ordered
configuration basis, a nearest-neighbor hop changes the minimum exactly when it moves the first
ordered particle. Holding the remaining $N-1$ coordinates fixed, the positions available to
that particle form an open chain, whose hopping block has norm at most $2J$. Summing directly
over all the fixed configurations therefore gives
\begin{equation}
 |Q_{\mathrm{meas}}|\le2J.
 \label{eq:energy_bound}
\end{equation}

The bound holds independently of the particle number, and independently of whatever quantum
correlations the state carries. It also survives any
interaction that is diagonal in position, since such a term commutes with the measurement and
therefore survives the subtraction unchanged. On a finite lattice of $M$ sites, which is
the geometry we simulate, the exact norm is $2J\cos[\pi/(M-N+2)]$. Resolving $r$ ordered
coordinates rather than one gives $|Q_{\mathrm{meas}}|\le2rJ$, and Appendix~\ref{app:energy}
proves all of this directly in the ordered occupation basis.

In a stationary engine, Eq.~\eqref{eq:energy_bound} becomes considerably more restrictive than
it first appears. Combining it with the balance of Eq.~\eqref{eq:stationary},
\begin{equation}
 0\le E\le2J,\qquad
 \langle m\rangle\le\frac{2}{\alpha N}.
 \label{eq:mean_bound}
\end{equation}
However many particles are present, the measurement can supply at most $2J$ per cycle, whereas a wall displacement $m$ has to lift
all $N$ of them through $\Delta m$ each. Adding particles therefore forces the mean advance
$\langle m\rangle$ down, simply because the same finite energy has to be shared out more widely.

That ceiling on $\langle m\rangle$ turns into a ceiling on the record. A nonnegative integer
variable whose mean is $\mu$ can have entropy at most
\begin{equation}
 g(\mu)=(1+\mu)\ln(1+\mu)-\mu\ln\mu,
 \label{eq:g}
\end{equation}
a value attained by the geometric distribution of that mean. Setting $\mu=\langle m\rangle$,
and using the fact that an entropy rate never exceeds the corresponding single-cycle marginal
entropy, Eq.~\eqref{eq:mean_bound} gives
\begin{equation}
 H_{\min}\le g\!\left(\frac{E}{\Delta N}\right)
          \le g\!\left(\frac{2}{\alpha N}\right).
 \label{eq:stationary_entropy}
\end{equation}
Equation~\eqref{eq:stationary_entropy} says more than that the record stays bounded as
particles are added, since its right-hand side actually decreases, falling asymptotically as
$[2/(\alpha N)]\{1+\ln(\alpha N/2)\}$ at fixed positive tilt. The inequality applies to every
stationary state with the finite internal means that Eq.~\eqref{eq:stationary} requires. It
does not, however, assert that the measured entropy is monotonic in $N$ at every finite
particle number.

\begin{figure}[tbp]
 \centering
 \includegraphics[width=\columnwidth]{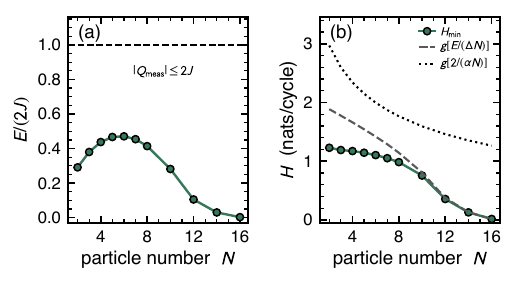}
 \caption{%
    A boundary measurement supplies only a finite amount of energy even as the particle number
    grows. (a)~Stationary stored work at $\alpha=0.15$ and $\tau=1.85$. The dashed line is the
    exact bound $2J$. (b)~The corresponding record entropy rate. The dashed curve is the ceiling
    obtained from the measured work, $g[E/(\Delta N)]$, and the dotted curve is the
    general stationary ceiling $g[2/(\alpha N)]$. The numerical points are stationary
    averages of the OS demon.}
 \label{fig:bounds}
\end{figure}

Figure~\ref{fig:bounds} shows how these two constraints appear in the noninteracting engine.
The stored work in Fig.~\ref{fig:bounds}(a) never reaches even half of $2J$, rising to $0.47$
of the bound at $N=6$ and falling away on either side. The measurement therefore supplies its
largest useful energy at intermediate filling rather than at the largest particle number, and
the decline above $N=6$ is the same suppression that Sec.~\ref{sec:scaling} later resolves
microscopically. These points, and every numerical result that follows, come from stationary
trajectories whose lengths, system sizes, and convergence checks are collected in
Appendix~\ref{app:numerics}.

The two ceilings in Fig.~\ref{fig:bounds}(b) behave differently, and the difference identifies
what controls the record. The measured entropy approaches the ceiling built from the stationary
work as particles are added, reaching $0.65$ of it at $N=2$ and $0.98$ at $N=14$. Beyond
$N\simeq10$ the two are indistinguishable on the scale of the figure, so at large filling the
stationary work fixes the record cost to within a few percent. The
ceiling built from the $2J$ bound alone falls only from $2.97$ to $1.35$ nats over the same
range, because it uses no information about the stationary work. The gap between the two
curves is therefore the part of the constraint that the dynamics supplies, as opposed to the
part that follows from the operator norm alone. The same argument also limits how often the
wall moves at all, since $\Pr(m\ge1)\le\langle m\rangle$ gives
\begin{equation}
 1-q_N\le\langle m\rangle\le\frac{2}{\alpha N},
 \qquad q_N=\Pr(m=0).
 \label{eq:stall_bound}
\end{equation}
Large filling therefore forces the no-advance probability toward unity even before the detailed
stationary state is known.

\subsection{Bound on the minimum record}

The bound just derived rests on the energy balance of the repeated engine, and so applies to
the stationary state. A second bound holds cycle by cycle, for any state of $N$ fermions
whatever, correlated or not. Let
a fixed-$N$ fermionic state satisfy the occupied-origin condition
$\hat n_0\rho=\rho$, and let $U$ be a quadratic number-conserving single-particle propagator.
Then
\begin{equation}
 \Pr(m\ge r)\le\sum_{j\ge r}|U_{j0}|^2.
 \label{eq:tail}
\end{equation}
Occupying the origin singles out one particle, and quadratic evolution carries that particle,
and only that particle, into the orbital $U|0\rangle$. Pushing every
particle past site $r$ therefore costs at least the weight that this one orbital has beyond
$r$, which is the right-hand side of Eq.~\eqref{eq:tail}. Appendix~\ref{app:tail} gives the
Fock-space proof and extends it to mixed states.

Summing the tails gives $\langle m\rangle\le\mu_1$, where
$\mu_1=\sum_jj|U_{j0}|^2$ is the displacement of one particle initially at the wall. Energy
conservation and $\|K\|\le2J$ give $\mu_1\le2/\alpha$. Hence for every conditioned history,
\begin{equation}
 H(m\mid\mathrm{history})\le g(\mu_1)\le g(2/\alpha).
 \label{eq:transient_entropy}
\end{equation}
The distinction is useful. Occupation of the origin is sufficient to bound the OS
record after quadratic evolution. Gaussian closure is needed only to propagate the complete
conditioned many-particle dynamics exactly over repeated cycles.

\section{Record cost and collective advantage}
\label{sec:advantage}

The FC demon is the natural point of comparison, since at the same tilt and cycle time
it applies exactly the same barrier rule and differs only in how much it resolves. Its record,
however, grows rapidly with $N$ [Fig.~\ref{fig:setup}(c)], and the reason that extra
information buys so little transport is a matter of geometry. How much work a particle
contributes depends on how far from the wall it starts, and that contribution collapses once
the starting site lies beyond the distance the dynamics can cover in a single cycle.
Appendix~\ref{app:kernel} makes this statement precise, and proves it with a Duhamel estimate
that runs over the whole propagation history.

Particles far from the wall therefore add many distinctions to a full configuration record
while contributing almost nothing to the work that cycle produces. The resulting imbalance is
already visible at matched cycle time, where Table~\ref{tab:matched} sets the FC entropy
and work against the single-particle values at $\alpha=0.3$. The record cost stays close to
additive, whereas the work is strongly sub-additive. At $\tau=2.5$, for instance, adding four
particles multiplies the entropy cost by about $4.3$ but the work by only about $1.4$, and the
two scalings separate further still as the cycle time grows.

\begin{table*}[tb]
  \caption{The FC demon pays an almost additive record for a strongly sub-additive
  work and does not reach the single-particle reference. The first two blocks give entropy
  and work per particle at matched cycle time, relative to one particle, at $\alpha=0.3$.
  The third block gives the ratio to the reference after optimizing the cycle time, at
  $\alpha=0.3$ and $0.5$.}
  \label{tab:matched}
  \begin{ruledtabular}
  \begin{tabular}{lccccccc}
    & $N=2$ & $N=3$ & $N=4$ & $N=5$ & $N=6$ & $N=7$ & $N=8$ \\ \hline
    \multicolumn{8}{l}{$H_{\mathrm{full}}(N)/[N H_1]$} \\
    $\tau=2.5$ & 0.976 & 0.929 & 0.896 & 0.870 & 0.857 & 0.844 & 0.834 \\
    $\tau=4.5$ & 0.949 & 0.904 & 0.877 & 0.853 & 0.837 & 0.820 & 0.808 \\
    $\tau=6.5$ & 0.925 & 0.885 & 0.860 & 0.841 & 0.826 & 0.811 & 0.800 \\
    \hline
    \multicolumn{8}{l}{$E_N/E_1$} \\
    $\tau=2.5$ & 1.196 & 1.307 & 1.380 & 1.426 & 1.481 & 1.512 & 1.536 \\
    $\tau=4.5$ & 1.300 & 1.476 & 1.608 & 1.718 & 1.798 & 1.860 & 1.949 \\
    $\tau=6.5$ & 1.397 & 1.652 & 1.842 & 1.992 & 2.130 & 2.233 & 2.353 \\
    \hline
    \multicolumn{8}{l}{$R_N/R_{\mathrm{ref}}$, optimized over cycle time} \\
    $\alpha=0.3$ & 0.724 & 0.584 & 0.501 & 0.442 & 0.399 & 0.366 & 0.344 \\
    $\alpha=0.5$ & 0.705 & 0.571 & 0.486 & 0.429 & 0.385 & 0.349 & 0.326
  \end{tabular}
  \end{ruledtabular}
\end{table*}

The same separation is visible in Fig.~\ref{fig:setup}(c). At $\alpha=0.15$ and $\tau=1.85$
the FC entropy rises nearly linearly over the displayed range, while the OS
record stays near one nat and decreases as the minimum becomes concentrated near the wall.
How much of the FC record is irrelevant to the barrier can be measured on a single
history, without setting two different stationary processes against each other. On an FC
history we therefore define
\begin{equation}
 I_{\mathrm{unused}}=H(\boldsymbol n_t\mid\boldsymbol n_{<t})
                   -H(m_t\mid\boldsymbol n_{<t}).
 \label{eq:unused}
\end{equation}
Both terms refer to the same conditioned state and to the same past, so their difference is
the information the demon writes down and never uses. Table~\ref{tab:waste} shows how that
unused fraction grows with particle number, rising from about $0.52$ at $N=2$ to $0.949$ at
$N=10$. By ten particles, then, almost the whole configuration record resolves distinctions
that never change where the barrier goes.

\begin{table}[tb]
  \caption{Unused information in the full-configuration record at $\alpha=0.15$ and
  $\tau=1.85$. $H_m^{(\mathrm{fc})}$ is the entropy rate of the minimum evaluated on the same
  full-configuration history as $H_{\mathrm{full}}$. The difference is therefore information in the
  configuration record that is not required to choose the barrier. Values for $N\ge2$ average
  four independent $4\times10^3$-cycle trajectories. Parentheses give the half-range across
  those trajectories in the final digit(s). $N=1$ is exact.}
  \label{tab:waste}
  \begin{ruledtabular}
  \begin{tabular}{lcccc}
    $N$ & $H_m^{(\mathrm{fc})}$ & $H_{\mathrm{full}}$ & $I_{\mathrm{unused}}$ & $I_{\mathrm{unused}}/H_{\mathrm{full}}$ \\ \hline
     1 & 1.309       &  1.309    &  0.000    & 0.000      \\
     2 & 1.2432(31)  &  2.580(20)&  1.337(18)& 0.5181(31) \\
     4 & 0.9441(97)  &  4.623(61)&  3.679(51)& 0.79576(91)\\
     6 & 0.749(12)   &  6.586(80)&  5.838(70)& 0.88634(67)\\
     8 & 0.614(15)   &  8.57(20) &  7.96(18) & 0.92833(66)\\
    10 & 0.5351(74)  & 10.54(17) & 10.01(16)& 0.94924(62)
  \end{tabular}
  \end{ruledtabular}
\end{table}

The discarded distinctions can still change the post-measurement state and later motion. The
OS demon must therefore be propagated on its own conditioned histories rather than
inferred by subtracting the stationary record costs of two different demons. The comparison in
Eq.~\eqref{eq:unused} diagnoses the information mismatch on one history. Whether removing that
mismatch can coexist with a collective work response is the question the rest of this section
answers.

Our main operating point for the OS demon is
\begin{equation}
 \alpha=0.15,\qquad N=7,\qquad \tau=1.85.
 \label{eq:headline_point}
\end{equation}
At this operating point the engine stores $E=\HeadlineE J$ per cycle while recording
$H_{\min}=\HeadlineH$ nats, which amounts to $R=\HeadlineR J$ of work per recorded nat at a
power per particle of $\mathcal P=\HeadlineP J^2/\hbar$. Measured against the full single-particle
envelope of Eq.~\eqref{eq:reference}, this gives
\begin{equation}
 \frac{R_N}{R_{\mathrm{ref}}}=\HeadlineRatio\pm\HeadlineSpread,
 \qquad N=7,
 \label{eq:headline}
\end{equation}
where the quoted figure after each value is the half-range across four independent long
trajectories, not a standard error. Toward
weaker tilt the gain grows further still, reaching $1.569\pm0.011$ at $\alpha=0.10$ with
$N=10$ and $\tau=1.85$.

Two quantities already established above combine to produce this ratio. The record cost obeys
Eq.~\eqref{eq:stationary_entropy} and stays near one nat, and the stored work is
$E=\Delta N\langle m\rangle$ by Eq.~\eqref{eq:wall_work}, so one wall advance raises all $N$
particles rather than one. Independent single-particle engines have neither. Each pays a
separate record for a separate particle, and each decision moves that particle alone. The
advantage is the separation between an information cost fixed by a boundary measurement and a
displacement shared by the whole cloud.

\section{Limits set by Pauli blocking}
\label{sec:scaling}

The advantage cannot keep growing with particle number. Adding
particles at first lets one barrier decision displace a larger cloud while the minimum record
stays just as compact, which is the whole source of the gain. Past a certain filling, however,
the site just above the wall is increasingly likely to be occupied again once the free
evolution ends. The wall then advances less and less often, and both the work and the record
activity collapse together.

\begin{figure}[tbp]
 \centering
 \includegraphics[width=\columnwidth]{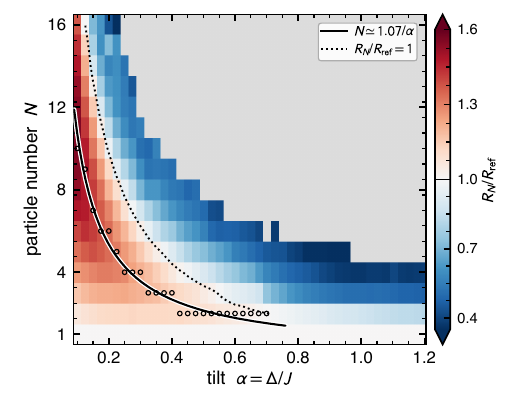}
 \caption{%
    The collective advantage occupies an intermediate particle-number range. Colour gives the
    maximum $R_N/R_{\mathrm{ref}}$ found over the cycle times considered. The dotted contour
    marks unit ratio. Black points mark the best particle number found in the scan, and the
    solid curve $N\simeq1.07/\alpha$ is an empirical guide through the collective region. Grey
    cells indicate points where wall motion is too rare to resolve reliably. $N=1$ is the
    optimized single-particle control.}
 \label{fig:gain_map}
\end{figure}

Figure~\ref{fig:gain_map} shows the region of collective gain that results. At each tilt we
vary the cycle time and keep the largest ratio found, and the particle number that maximizes
the ratio follows
an approximate $1/\alpha$ trend across the whole region. A least-squares guide through the
coarse-grid maxima gives $N\simeq1.07/\alpha$, though neighboring integers are not always
resolved from one another. At $\alpha=0.30$, for example, refined values for $N=3$ and $4$ are
$1.2205\pm0.0050$ and $1.2201\pm0.0067$, which are indistinguishable within their trajectory
spreads.

\begin{table}[tb]
  \caption{Operating points of the OS demon at the particle number the broad scan selects,
   with the cycle time and the ratio then refined. $N_{\mathrm{best}}$ is the particle number
   with the largest ratio in the scan, $\tau^*$ is the cycle time that maximizes the ratio at
   that $N$ on a grid of $0.025$, and $z(\tau^*)$ is the single-particle Bloch excursion of
   Eq.~\eqref{eq:bloch}. Each row uses four independent $1.5\times10^4$-cycle trajectories, and
   the figure after each ratio is their half-range. Repeating every row on $M=280$ changes no
   digit shown. Neighboring integer particle numbers can be nearly degenerate, so
   $N_{\mathrm{best}}$ is not resolved from its neighbors at every tilt.}
   \label{tab:peaks}
   \begin{ruledtabular}
   \begin{tabular}{lccccc}
     $\alpha$ & $N_{\mathrm{best}}$ & $\tau^*$ & $R_N/R_{\mathrm{ref}}$ & $R_N/J$ & $z(\tau^*)$ \\ \hline
    0.10 & 10 & 1.850 & $1.569\pm0.011$ & 0.834 & 3.69 \\
    0.15 &  7 & 1.850 & $1.440\pm0.011$ & 0.860 & 3.69 \\
    0.20 &  6 & 1.750 & $1.334\pm0.006$ & 0.871 & 3.48 \\
    0.25 &  4 & 1.900 & $1.281\pm0.010$ & 0.900 & 3.76 \\
    0.30 &  4 & 1.725 & $1.220\pm0.007$ & 0.912 & 3.41 \\
    0.40 &  3 & 1.750 & $1.144\pm0.006$ & 0.946 & 3.43 \\
    0.50 &  2 & 1.900 & $1.099\pm0.002$ & 0.982 & 3.66 \\
    0.60 &  2 & 1.725 & $1.059\pm0.005$ & 1.010 & 3.30 \\
    0.70 &  2 & 1.650 & $1.011\pm0.003$ & 1.020 & 3.12
   \end{tabular}
   \end{ruledtabular}
 \end{table}

Table~\ref{tab:peaks} makes two features explicit. First, the relative gain is largest at weak
tilt, where the optimized single-particle engine is least economical with the information it
records. The absolute work per nat nevertheless stays substantial right across the collective
region. Second, the excursion $z(\tau^*)$ varies far less than $1/\alpha$ does, which tells us
that the distance explored within one cycle and the number of particles the stationary cloud
can hold are set by different things.

The microscopic signature of the suppression is the no-advance probability
\begin{equation}
 q_N=\Pr(m=0)=(U\Gamma U^\dagger)_{00}.
 \label{eq:q}
\end{equation}
This identity is exact, because the event $m=0$ is nothing other than occupation of site zero
once the free evolution has ended. For a Fock state with the particles sitting at the sites $n_a$, it
reduces to the single-particle sum $q_N=\sum_a|U_{0n_a}|^2$. A general conditioned state,
however, also carries the off-diagonal elements of $\Gamma$, and those coherent contributions
are large enough numerically that replacing $q_N$ by the single-particle sum would be wrong.

The natural spatial scale is set by Bloch motion. On the unbounded tilted lattice a particle
localized at one site has a Bessel-distributed displacement with argument
\begin{equation}
 z(\tau)=\frac{4}{\alpha}\left|\sin\frac{\alpha\tau}{2}\right|.
 \label{eq:bloch}
\end{equation}
Its maximum $4/\alpha$ is the Bloch-excursion scale
\cite{Bloch1929,Wannier1960,Gluck2002,Hartmann2004}. The stationary many-particle cloud is not a
compact wave packet of this width, so $z(\tau)$ is a characteristic excursion scale rather than
a strict support bound.

With this interpretation, $z(\tau)$ sets the scale for how far the boundary can reach
occupied sites within one cycle. Along the coarse-grid collective maxima for $0.1\le\alpha\le0.7$, the
optimal $z(\tau^*)$ has mean $3.55$ and standard deviation $0.25$. The cycle therefore uses a
similar short-time excursion while the stationary number of particles that the feedback can lift grows
roughly as $1/\alpha$.

\begin{figure}[tbp]
 \centering
 \includegraphics[width=\columnwidth]{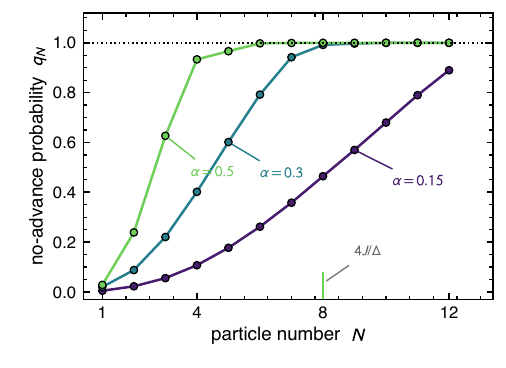}
 \caption{%
    Pauli blocking suppresses wall motion as the accessible region fills. The no-advance
    probability $q_N$ is shown against particle number for three representative tilts. The
    dotted line marks complete blocking, $q_N=1$. The short mark on the horizontal axis indicates
    the Bloch-excursion scale $4J/\Delta$ for $\alpha=0.5$.}
 \label{fig:stall}
\end{figure}

Figure~\ref{fig:stall} makes the approach to blocking explicit. As the particle number grows,
$q_N$ approaches one and the minimum rarely leaves the origin. The exact inequality in
Eq.~\eqref{eq:stall_bound} already forces this trend at large $N$. Once the advances become very
rare, assigning a unique integer stall point depends on how long they are observed. We therefore
use the continuous growth of $q_N$ as the microscopic signature of blocking. The useful
collective regime lies at intermediate filling, before Pauli exclusion suppresses both wall
motion and record production. Strongly tilted lattices are an active experimental setting in
their own right, where constrained hopping slows relaxation and fragments the accessible
Hilbert space \cite{Scherg2021,GuardadoSanchez2020,Kohlert2023,Sala2020}.

This also clarifies the role of the particle-number optimum. It is a dynamical compromise, not a
second entropy theorem. At small $N$ the wall decision is inexpensive but acts on too few
particles. At large $N$ the measurement still has a bounded energy input, but the occupied
boundary prevents it from translating the cloud. The maximum in Fig.~\ref{fig:gain_map} lies
between these two limits.

\section{Information saving and retained coherence}
\label{sec:coherence}

Measurement resolution governs not only what the demon records, but also the quantum state it
leaves for the next cycle. An FC position measurement destroys the coherence between
configurations that call for the same barrier, whereas the OS measurement leaves
that whole sector unresolved and the coherence inside it intact. Both consequences follow from
the same choice of measurement, yet they act on the engine in different ways, and the point of
this section is to tell them apart.

\subsection{One-body coherence}

Coherence enters the work through the off-diagonal part of the one-body density matrix, which
the decomposition below isolates exactly. Let $\Gamma$ be evaluated immediately after feedback, and let
$U$ be the single-particle propagator over one free evolution. Write $x$ for the
single-particle position operator, diagonal in the site basis with the site index as its
eigenvalue. The energy a particle gains over
that evolution is then measured by
\begin{equation}
 W=\Delta(U^\dagger xU-x),
 \label{eq:work_operator}
\end{equation}
which is simply $\Delta$ times the displacement operator. In terms of $W$, the work of the next
unitary step splits exactly into a diagonal and an off-diagonal part,
\begin{align}
 E_{\mathrm{unitary}}&=\mathrm{Tr}(W\Gamma),&
 E_{\mathrm{diag}}&=\mathrm{Tr}[W\,\mathrm{diag}(\Gamma)],\nonumber\\
 E_{\mathrm{coh}}&=E_{\mathrm{unitary}}-E_{\mathrm{diag}}.
 \label{eq:coherence}
\end{align}
where $\mathrm{diag}(\Gamma)$ keeps only the site populations. The term
$E_{\mathrm{diag}}$ is thus the work the same populations would give if all coherence were
destroyed, and $E_{\mathrm{coh}}$ is what the surviving coherence adds on top. All three are
evaluated on the same conditioned state, before the next outcome is drawn, so the decomposition
carries no measurement shot noise and needs no endpoint correction. Averaged over the
stationary cycle, $E_{\mathrm{unitary}}$ agrees with the stored work of
Eq.~\eqref{eq:stationary}.

At the seven-particle operating point we find $E_{\mathrm{diag}}=\HeadlineDiag J$ and
$E_{\mathrm{coh}}=\HeadlineCoh J$, so the coherent term accounts for about
$\HeadlineCohPercent\%$ of the mean unitary work. For the FC demon, by contrast, the
post-feedback state is diagonal in the site basis, and $E_{\mathrm{coh}}$ therefore vanishes in
every single cycle rather than merely on average.

\begin{figure}[tbp]
 \centering
 \includegraphics[width=\columnwidth]{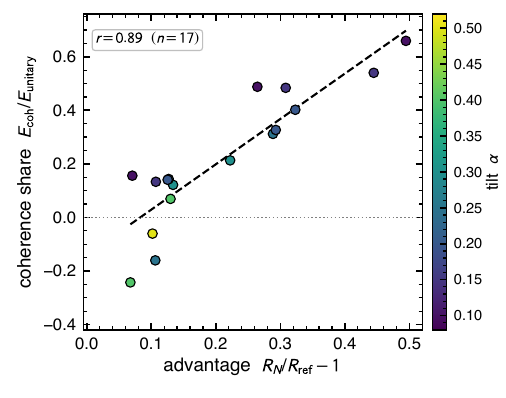}
 \caption{%
    Coherent work contribution against the collective advantage. Points give the exact
    coherence share $E_{\mathrm{coh}}/E_{\mathrm{unitary}}$ for the OS demon,
    coloured by tilt, at operating points with $R_N/R_{\mathrm{ref}}\ge0.95$. The dashed line
    is a least-squares fit. For the FC demon the same quantity vanishes identically
    because the post-measurement one-body density matrix is diagonal in the site basis.}
 \label{fig:coherence}
\end{figure}

Figure~\ref{fig:coherence} shows that the coherent share grows with the collective advantage
over these operating points, with a correlation coefficient close to $0.9$. The contribution is
positive throughout most of the weak-tilt regime and can become negative at larger filling.
The sign change belongs to the conditioned transport state rather than to a universal notion of
coherence as a resource \cite{Baumgratz2014,Streltsov2017}. What coherence can contribute to
work is constrained by more than the free energy alone
\cite{Lostaglio2015,Korzekwa2016,Francica2019}.

\subsection{Separating coherence from information saving}

A stronger test is to remove the coherence while keeping the same coarse record. After measuring
$m$, apply complete position dephasing without recording which configuration occurred,
\begin{equation}
 \mathcal D(\rho)=\sum_{\boldsymbol n}|\boldsymbol n\rangle\langle\boldsymbol n|
                  \rho|\boldsymbol n\rangle\langle\boldsymbol n|.
 \label{eq:dephasing}
\end{equation}
Such dephasing leaves the recorded variable untouched, while changing both the state and the
energy the measurement supplies. For $N=2$ the resulting wall-frame state is a classical
mixture over the separation between the two particles. We evolve that mixture conditioned only
on the observed sequence of minima, so that the demon never learns the hidden configuration. Appendix~\ref{app:dephasing} gives the exact update.

Once the extra dephasing is applied, the OS demon has the same unconditional
configuration dynamics as an FC measurement whose detailed record is thrown away. From a
given initial ensemble, its mean work is therefore identical to the FC value. What
still differs between them is the record, since one demon stores only $m$ while the other
stores the whole configuration. Comparing those two cases thus isolates the information saving
by itself. Comparing either of them with the L\"uders OS demon brings in the
retained coherence as well.

\begin{table}[tbp]
 \caption{Two-particle comparison at $\alpha=0.15$. Energy is in units of $J$ and entropy in
 nats per cycle. Values are obtained from two independent trajectories. The last column uses
the single-particle reference at the power of the same row.}
 \label{tab:dephasing}
 \begin{ruledtabular}
 \begin{tabular}{clrrr}
 $\tau$ & Measurement & $E$ & $H$ & $R/R_{\mathrm{ref}}$\\
 \hline
 4.4 & Order statistic & 1.371 & 2.073 & 1.107\\
     & Dephased order statistic & 1.163 & 2.201 & 0.885\\
     & Rank one & 1.158 & 3.892 & 0.498\\
 8.0 & Order statistic & 1.499 & 2.432 & 1.032\\
     & Dephased order statistic & 2.061 & 2.753 & 1.254\\
     & Rank one & 2.054 & 4.985 & 0.690
 \end{tabular}
 \end{ruledtabular}
\end{table}

The two operating points in Table~\ref{tab:dephasing} separate the mechanisms. At
$\tau=4.4$, the L\"uders OS measurement retains coherence and gives more work than the
dephased OS demon, lifting the engine above the single-particle reference. At $\tau=8$, the
fully dephased OS demon itself reaches $R/R_{\mathrm{ref}}=1.254$, showing that the reduced
record cost can support a collective advantage even when the post-measurement state is
classical in position.

Complete dephasing also changes the measurement backaction by removing additional hopping
coherences and can supply energy of its own. At $\tau=8$ this additional backaction raises the
work above the L\"uders value. The $2J$ bound of Eq.~\eqref{eq:energy_bound} applies to the
L\"uders OS measurement alone, whereas complete position dephasing touches the coherences of
all $N$ ordered coordinates and therefore falls under the $r=N$ version of the same bound.
Together, the three rows separate the reduction in recorded information from the energetic and
coherent backaction of the measurement that produces the record.

% ============================================================================
\section{Errors in barrier placement}
\label{sec:errors}

An ideal demon moves the wall by exactly the measured displacement in every cycle. To test what
a placement error costs, without at the same time changing what the demon knows, we let the
order statistic be measured correctly and record separately whether the wall was misplaced.
Specifically, for $m\ge1$ the wall advances by $m-1$ rather than by $m$ with probability
$\epsilon$, while $m=0$ leaves it where it is. Since that binary outcome is itself written to
the tape, the entropy rate of the record becomes
\begin{equation}
 H_{\mathrm{tape}}=H(m_t\mid\mathrm{past})+
       \Pr(m_t\ge1)h_2(\epsilon),
 \label{eq:error_record}
\end{equation}
where $h_2$ is the binary entropy in nats.

Table~\ref{tab:imprecise} gives the stationary performance that follows. The work falls as
placement errors accumulate, while the tape entropy rises for two reasons. The misplacement is
itself a recorded variable, and earlier errors alter the state handed on to later cycles. The ratio nevertheless stays above the single-particle reference
through $\epsilon=0.10$, and only by $\epsilon=0.20$ does it drop below.

\begin{table}[tb]
  \caption{Effect of a monitored one-site under-placement at the main operating point
  $\alpha=0.15$, $N=7$, and $\tau=1.85$. $H_{\mathrm{tape}}$ includes the minimum record and
  the recorded misplacement. Values are obtained from four independent
  $1.5\times10^4$-cycle trajectories. Parentheses in the last column give the half-range of
  the trajectory ratios in the final digit(s).}
  \label{tab:imprecise}
  \begin{ruledtabular}
  \begin{tabular}{lccc}
    $\epsilon$ & $E/J$ & $H_{\mathrm{tape}}$ & $R/R_{\mathrm{ref}}$ \\ \hline
    0.00 & 0.905 & 1.053 & 1.440(11) \\
    0.02 & 0.886 & 1.125 & 1.319(22) \\
    0.05 & 0.861 & 1.193 & 1.208(8)  \\
    0.10 & 0.817 & 1.276 & 1.073(10) \\
    0.20 & 0.743 & 1.385 & 0.898(8)  \\
    0.30 & 0.683 & 1.444 & 0.792(4)
  \end{tabular}
  \end{ruledtabular}
\end{table}

Table~\ref{tab:imprecise} characterizes a monitored one-site under-placement error, with the
misplacement included in the recorded tape.

\section{Experimental implementation}
\label{sec:experiment}

The cycle asks for the same basic cold-atom elements as the single-particle engine, together
with two further requirements that are specific to the many-particle case. First, the
measurement must locate the leftmost atom without resolving the rest of the configuration.
Second, the phase that accumulates while the lattice is frozen has to be controlled. We
therefore work through a concrete $^6$Li realization in which both requirements are kept
explicit.

\subsection{Platform and operating point}

The estimates below refer to a concrete set of lattice parameters, fixed here.
A vertical $^6$Li lattice at wavelength $1064$ nm has spacing $d=532$ nm, and gravity then
supplies the step $\Delta/h=78.64$ Hz. To reach $\alpha\simeq0.15$, a plane-wave band
calculation puts the operating depth at $10.30E_r$, where $E_r$ is the lattice recoil energy
with $E_r/h=29.30$ kHz, and gives $J/h\simeq524$ Hz. At that depth the gap to the next band is
$137$ kHz, comfortably larger than every other scale in the problem, and the free evolution of
Eq.~\eqref{eq:headline_point} lasts $0.562$ ms. The same band structure gives a
next-nearest-neighbor hopping of about $1.1\%$ of the nearest-neighbor value, which is a direct
estimate of the leading correction to the ideal Hamiltonian.

\subsection{Preparation}

A single-spin-component gas can be prepared as a unit-filled band-insulating segment and imaged
with established fermionic quantum-gas microscopy techniques
\cite{Parsons2015,Cheuk2015,Haller2015,GrossBakr2021}. Site-resolved imaging and single-site
addressing are by now routine in optical lattices
\cite{Bakr2009,Sherson2010,Weitenberg2011,Simon2011,Fukuhara2013,GrossBloch2017}, and the same
microscopes resolve spin and density in fermionic chains
\cite{Boll2016,Mazurenko2017}. Pauli blocking itself has been observed directly in degenerate
fermionic lattice gases \cite{Omran2015}. Faster imaging, and imaging below the lattice spacing, shortens the interrogation each cycle
needs \cite{Su2025,Alberti2016}. The experiment begins from this prepared Fock state, and after
the feedback cycles are complete, a final site-resolved image gives the laboratory-frame
potential energy through Eq.~\eqref{eq:finite_work}.

\subsection{Frozen interrogation}

The minimum can be interrogated while hopping is suppressed, which is done by raising the
lattice depth. That freezes the spatial occupations, but it does not freeze the tilt, and over
a hold of duration $t_{\mathrm{fr}}$ the state therefore acquires
\begin{equation}
 D_\phi=e^{-i\phi\hat X},\qquad \phi=\Delta t_{\mathrm{fr}}/\hbar.
 \label{eq:phase}
\end{equation}
For a $360\,\mu$s frozen interval at the operating point, this amounts to $0.178$ rad per site.
A configuration eigenstate would simply pick up an overall phase and be none the worse for it.
The OS measurement, however, retains superpositions of configurations within one
sector, so the relative phase between them changes the coherent evolution that follows.

The phase can be removed without changing the measurement outcome. Consider a cycle consisting
of free evolution, a rapid intraband ramp to frozen hopping, minimum interrogation, wall
feedback, an inverse phase imprint, and restoration of the operating lattice. In the ideal
frozen limit the conditional operator is
\begin{equation}
 A_m=D_{-\phi}T_mP_mD_\phi\mathcal U_\tau
       =e^{-i\phi Nm}T_mP_m\mathcal U_\tau.
 \label{eq:compensated_cycle}
\end{equation}
Since the last factor is only an outcome-dependent global phase, the compensated cycle
reproduces Eq.~\eqref{eq:cycle} exactly. One could equally compensate the force itself during
the interrogation interval. Either way, if the number of threshold questions varies from cycle
to cycle, the correction has to be built from the hold duration actually used.

This compensation is essential rather than cosmetic. Keeping the ideal free evolution fixed and
adding an uncompensated residual phase of $0.05$, $0.10$, and $0.178$ rad per site per cycle
changes the seven-particle ratio to $1.217$, $1.017$, and $0.746$ respectively. The last of
these already destroys the advantage. The ideal result is therefore a property of a coherent
cycle, and not one that survives an arbitrary frozen hold.

When the frozen interval is implemented by changing the lattice depth, the ramps also enter the
energy balance. At exactly frozen hopping, $H_{\mathrm{fr}}=\Delta X$ commutes with the position
measurement, so the readout itself has no mean energy backaction. The ramp work is
\begin{equation}
 W_{\mathrm{ramp}}=\int dt\,\dot J(t)
             \left\langle\frac{\partial H}{\partial J}\right\rangle.
 \label{eq:ramp_work}
\end{equation}
In the ideal compensated limit this work supplies the kinetic-energy change assigned to the
instantaneous measurement in the theoretical cycle. The experimental energy balance must
therefore include the lattice ramps together with the probe and the phase-compensation step.

\subsection{Measurement of the order statistic}

The demon has to find the leftmost atom without learning anything else, which is the most
demanding requirement the protocol places on the apparatus. An adaptive threshold search can
locate $m$ using the projectors
\begin{equation}
 \begin{aligned}
 Q_{\mathrm{empty}}(s)&=\prod_{j=0}^{s}(1-\hat n_j),\\
 Q_{\mathrm{occupied}}(s)&=I-Q_{\mathrm{empty}}(s).
 \end{aligned}
 \label{eq:threshold}
\end{equation}
Here $Q_{\mathrm{empty}}(s)$ asks whether every site up to $s$ is empty, and a nested sequence
of such binary answers multiplies out to the projector $P_m$. The sequence realizes the
OS measurement only if each step is close to the corresponding L\"uders
projection.
Once the binary outcome is fixed, the probe should reveal neither the number of marked atoms nor
their positions inside the marked region. Otherwise the environment distinguishes
configurations that the ideal measurement leaves coherent.

Dispersive cavity detection offers a route to the required occupied-or-empty response
\cite{Kroeze2023,Orsi2024,Shadmany2025,Shaw2026}. Existing experiments demonstrate high
cooperativity, lithium cavity control, and rapid single-atom detection. Cavities already read
atoms out nondestructively in the middle of a sequence \cite{Deist2022} and count them without
resolving which atom is which \cite{Hume2013}, which establishes the sensitivity and the
nondestructive character the threshold search needs. A number-resolved readout is not yet the
instrument assumed here, since the count itself separates configurations that share a minimum.
The probe must in addition be made insensitive to how many atoms the marked region holds.
The feedback must also close within a cycle, and lattice experiments now run
real-time control at that speed \cite{FPGA2026}. The remaining condition
is that the binary signal carry negligible configuration information within the accepted
sector. This is the distinction between detecting whether a region is occupied and realizing
the coarse measurement assumed in Eq.~\eqref{eq:cycle}.

For a known conditional distribution, the thresholds can be chosen to minimize the expected
number of questions. Appendix~\ref{app:experiment} gives the exact ordered-search recurrence.
At the main operating point the order statistic requires about two threshold questions on average.
The number of questions sets the probe exposure, while the record cost is still the entropy
rate in Eq.~\eqref{eq:entropy_rate}.

The same apparatus separates the two demons by how often it must ask. Resolving the full
configuration means producing the record whose cost Table~\ref{tab:waste} gives as $8.57$ nats
at $N=8$. Any protocol built from binary questions then needs at least twelve of them on
average, against about two for the order statistic. The question count therefore
distinguishes the two measurements operationally, without reference to the model.

\subsection{Backaction and heating}

Any probe that scatters light heats the sample, so we estimate what one scattering event costs.
Probe recoil is set by the actual $671$-nm momentum transfer rather than by the lattice
recoil. If the
axial momentum transfer is $\Delta k_z$ in a frozen well of frequency $\omega_{\mathrm{fr}}$, one
scattering event adds the harmonic-oscillator occupation
\begin{equation}
 \delta\bar n=
 \frac{\hbar\langle(\Delta k_z)^2\rangle}{2m_{\mathrm{at}}\omega_{\mathrm{fr}}}.
 \label{eq:recoil}
\end{equation}
The excitation probability then follows from this momentum transfer, the frozen-well frequency,
and the number of scattering events. Trap-laser noise heats the sample through a separate
channel and sets an independent limit on how long the cycle may run \cite{Savard1997}.

\subsection{Statistics and finite runs}

An experiment runs for finitely many cycles, and what it measures over that run differs from
the stationary value. A finite run begins from a prepared state and ends with a final position image. Its measured
work need not equal the stationary value after only a few cycles. In $512$ independent
ideal runs of $30$ cycles from a packed seven-particle state, terminal imaging gives mean stored
energy $1.068J$ per cycle and mean tape surprisal $1.002$ nats per cycle. Wall displacement
alone would give $0.962J$ per cycle because the internal cloud has not yet lost its endpoint
contribution. The finite-run comparison therefore uses Eq.~\eqref{eq:finite_work} on both
the collective and single-particle engines.

The work and the tape surprisal fluctuate together, so the uncertainty of their ratio is best
evaluated from the paired totals of each run. Appendix~\ref{app:experiment} gives the estimator,
including correlations between successive cycles. For the ideal $30$-cycle protocol the
relative run-to-run spread of the ratio is $0.124$. Preparation, detection, and calibration
uncertainties enter separately in an experiment.

% ============================================================================
\section{Discussion and outlook}
\label{sec:discussion}

The many-particle engine reveals a separation between the information contained in a quantum
state and the information required to control it. Measuring the full configuration resolves an
increasing number of distinctions, even though the wall feedback uses only the position of the
leftmost particle. The FC record therefore grows rapidly with particle number while much
of that information never changes the barrier action. Measuring the order statistic directly
removes this growing resolution cost, yet the same wall continues to act on the many-particle
cloud. At the calibrated operating point this produces a $44\%$ advantage in work per recorded
nat over independent single-particle engines at the same tilt and at the same or greater
power per particle.

The exact bounds explain why the minimum record stays so economical. Because the minimum is a
boundary coordinate, its L\"uders measurement disturbs only the hopping coherences tied to the
leftmost ordered particle, and the energy it supplies is therefore bounded independently of
$N$. In a stationary wall frame, that same finite energy has to raise every particle whenever
the wall moves, which forces both the mean advance and the entropy ceiling downward as the
filling increases. A separate Fock-space argument shows that the transient bound follows from
the occupied-origin condition for arbitrary correlated fermionic states. Gaussian closure
enables exact propagation of the repeated noninteracting dynamics.

Fermi statistics then decides how far the collective response can be pushed. At low filling a
single barrier decision acts on several particles at once. As the accessible region just above
the wall fills
up, however, Pauli blocking makes the first allowed site ever more likely to be occupied again,
and the wall moves less and less often. The result is a maximum at intermediate particle
number, beyond which both transport and record production are suppressed. That optimum follows
an approximate $1/\alpha$ trend over the collective region. The onset of very rare wall motion
is described more directly by the continuous no-advance probability than by a universal integer
stall point.

Measurement resolution also governs the state that enters the next cycle. The FC
measurement removes the coherence between configurations calling for the same feedback, whereas
the OS measurement leaves that sector unresolved. The one-body coherence it retains can
contribute substantially to the work at weak tilt. The dephasing comparison shows that the
reduced record cost can already support a collective advantage in a position-diagonal state,
while retained coherence provides an additional work contribution. These two consequences of
the coarse measurement can be separated by changing the measurement backaction while holding
the recorded variable fixed.

The experimental implementation makes both distinctions concrete. Freezing the hopping does
not freeze the phase that the tilt generates, so a repeated coherent cycle calls for phase
compensation, or for some equivalent modification of the propagator. Locating the leftmost atom
without destroying the useful coherence likewise calls for a threshold measurement, one that
reveals whether a region is occupied without resolving the configuration inside it. These
conditions provide a direct experimental test of how measurement resolution affects the
engine.

Interactions provide a natural extension. Translation-invariant density interactions between
spinless fermions preserve the measurement-energy bound while changing the stationary dynamics,
as summarized in Appendix~\ref{app:statistics}. Multicomponent fermions and soft-core bosons add
further interaction channels \cite{Preiss2015,Scherg2021}, and existing cold-atom transport
experiments provide settings in which the same feedback principle can be tested
\cite{Brown2019,Nichols2019,Vijayan2020}. The broader question is how information cost and
measurement backaction constrain collective transport once correlations arise from interactions
as well as quantum statistics.

\begin{acknowledgments}
We thank Masaya Nakagawa and Masahito Ueda for fruitful discussions, and Takeshi Fukuhara for raising the questions that inspired this work. This work was supported by the Scientific Research Start-up Foundation
of Xihua University (Grant No. Z241064).
\end{acknowledgments}

\section*{Data availability}
The data and the code that generates them are available upon reasonable request
from the first author.

\appendix

\section{Gaussian reduction and its validation}
\label{app:gaussian}

Because the Hamiltonian is quadratic, every conditioned Slater state can be
followed through its occupied one-particle orbitals. We collect the $N$
orthonormal orbitals in an $M\times N$ matrix $\Phi$, with one-body density
matrix $\Gamma=\Phi\Phi^\dagger$. If $A_m=\{0,\ldots,m-1\}$, the probability
that the first $m$ sites are empty is
\begin{equation}
 S_m=\det(I-\Gamma_{A_m}),\qquad p(m)=S_m-S_{m+1}.
 \label{eq:determinant}
\end{equation}
These are the standard determinantal occupation identities for free fermions
\cite{Macchi1975,Klich2003,Hough2006}. The same correlation matrix also carries
the reduced density matrices of a free-fermion state
\cite{Peschel2003,PeschelEisler2009}. The minimum projection removes orbital
components below $m$ and enforces occupation of site $m$, and both operations
preserve a Slater determinant. The full-configuration measurement preserves the Slater
form as well, because it collapses the state to occupied site orbitals. Both
demons can therefore be propagated exactly on $M\times N$ matrices even though
their conditioned histories differ.

\subsection{The order-statistic projection}

The Slater determinant $|\Phi\rangle=\phi_1\wedge\cdots\wedge\phi_N$
obeys
\begin{equation}
 \Pi_{S,N}|\Phi\rangle=(P_S\phi_1)\wedge\cdots\wedge(P_S\phi_N).
 \label{eq:wedge}
\end{equation}
Its squared norm is $\det(\Phi^\dagger P_S\Phi)$. Taking $S$ to exclude
the first $m$ sites gives
$\det(\Phi^\dagger P_S\Phi)=\det(I-\Gamma_{A_m})$ by the determinant
identity $\det(I-AB)=\det(I-BA)$.

Orthonormalize the surviving orbitals in Eq.~\eqref{eq:wedge}. Within their
occupied subspace, choose an orbital $\chi_1$ carrying the amplitude at site
$m$ and $N-1$ orbitals $\chi_2,\ldots,\chi_N$ with zero amplitude there.
For a nonzero-probability outcome,
\begin{equation}
 c_m|\Phi_S\rangle=\chi_1(m)\,
                    \chi_2\wedge\cdots\wedge\chi_N,
\end{equation}
up to the common determinant phase. The normalized state after occupation
of $m$ is therefore
\begin{equation}
 |m\rangle\wedge\chi_2\wedge\cdots\wedge\chi_N.
 \label{eq:slater_update}
\end{equation}
Zeroing rows below $m$, orthonormalizing, and forming the orthogonal complement
of the site-$m$ amplitude vector implements this map exactly. It preserves
all coherences allowed by the minimum sector.

\subsection{Rank-one sampling}

For an orthonormal orbital matrix, the probability of a configuration is
\begin{equation}
 p(\boldsymbol n)=|\det\Phi_{\boldsymbol n}|^2.
 \label{eq:dpp}
\end{equation}
The sequential determinantal sampler selects a site with probability
$\Gamma_{jj}/N$, projects the occupied subspace onto vectors vanishing at
that site, and repeats in the remaining rank. Each ordering of a sampled
configuration has probability $p(\boldsymbol n)/N!$. The unordered sample
thus obeys Eq.~\eqref{eq:dpp} \cite{Hough2006}. The recorded surprisal is
$-\ln p(\boldsymbol n)$, evaluated from the determinant of the original
propagated orbitals at the selected sites.

\subsection{Dephased order-statistic evolution}
\label{app:dephasing}

For two particles, complete position dephasing after minimum readout gives
the wall-frame state
\begin{equation}
 \rho_t=\sum_{d\ge1}w_t(d)|0,d\rangle\langle0,d|.
\end{equation}
The transition probability from separation $d$ to minimum $m$ and final
separation $d'$ is
\begin{equation}
 T_{m,d',d}=|U_{m0}U_{m+d',d}-U_{md}U_{m+d',0}|^2.
 \label{eq:pair_kernel}
\end{equation}
Conditional on the observed minimum alone,
\begin{align}
 p_t(m)&=\sum_{d,d'}T_{m,d',d}w_t(d),\nonumber\\
 w_{t+1}(d')&=\frac{\sum_dT_{m,d',d}w_t(d)}{p_t(m)}.
 \label{eq:filter}
\end{align}
The record cost is $-\ln p_t(m)$ and the conditional potential-energy gain is
\begin{equation}
 E_t=\Delta\left[2m+\sum_dd\,w_{t+1}(d)-\sum_dd\,w_t(d)\right].
\end{equation}
The conditional distribution retains the unresolved separation. Summing over $m$ gives the
same unconditional measurement-feedback map as a fully dephased position measurement whose
configuration record is discarded. The two dephased measurements therefore give the same
ensemble work at every cycle when they start from the same ensemble, while their record entropy
rates remain different because they condition on different observations.

\section{Exact bounds for the cycle}

\subsection{Measurement energy and stationary balance}
\label{app:energy}

Work first on $M$ sites in the fixed-$N$ sector with ordered basis
$|n_1,\ldots,n_N\rangle=c_{n_1}^\dagger\cdots c_{n_N}^\dagger|0\rangle$,
$0\le n_1<\cdots<n_N\le M-1$. For nearest-neighbor hopping, the matrix element
between configurations differing by one allowed hop is $-J$. The hop cannot
exchange the ordering of particles. Decompose
\begin{equation}
 K=\sum_{a=1}^N K_a,
 \label{eq:rank_hopping}
\end{equation}
where $K_a$ moves the $a$th ordered coordinate by one site when the destination
is empty. At fixed values of all other coordinates, the allowed values of
$n_a$ form an interval of $L$ sites, with $L\le M-N+1$. The operator on this
interval has eigenvalues $-2J\cos[k\pi/(L+1)]$, $k=1,\ldots,L$. Therefore
\begin{equation}
 \|K_a\|=2J\cos\frac{\pi}{M-N+2}<2J
 \label{eq:finite_norm}
\end{equation}
for $N<M$, with zero norm when $N=M$. Equality follows by placing the other
particles at the ends of the lattice, leaving the maximal interval for $n_a$.

Let $A$ contain $r$ ordered coordinates, and let $P_{\boldsymbol q}$ be the
joint projectors onto their values. Every matrix element of $K_a$ with
$a\in A$ joins different measurement sectors. Every matrix element with
$a\notin A$ remains within a sector. Hence
\begin{equation}
 K-\sum_{\boldsymbol q}P_{\boldsymbol q}KP_{\boldsymbol q}
   =\sum_{a\in A}K_a,
 \label{eq:resolved_hopping}
\end{equation}
and its norm is at most $2rJ$. Any position-diagonal term $V_{\mathrm{d}}$ commutes
with the projectors. For $H=K+V_{\mathrm{d}}$ and arbitrary normalized $\sigma$,
\begin{equation}
 Q_{\mathrm{meas}}=-\mathrm{Tr}\left[\sigma\sum_{a\in A}K_a\right],
 \qquad |Q_{\mathrm{meas}}|\le2rJ.
 \label{eq:general_bound}
\end{equation}

The order-statistic measurement is $A=\{1\}$, for which
Eq.~\eqref{eq:finite_norm} is the exact finite-system norm. The semi-infinite
bound follows directly from the norm bound for the open half-line blocks.
The result depends on the projectors that define the measurement, not only on its outcome
labels. Additional dephasing within a fixed minimum sector changes further hopping coherences
and therefore contributes separately to the energy change.

For minimum feedback, let $\rho'_m=P_m\sigma P_m/p_m$. Since all sites below
$m$ are empty in this state, translating by $m$ preserves the kinetic energy.
It also preserves translation-invariant diagonal interactions, while
\begin{equation}
 \mathrm{Tr}(X T_m\rho'_mT_m^\dagger)=\mathrm{Tr}(X\rho'_m)-Nm.
 \label{eq:translation}
\end{equation}
The unitary part preserves total energy. Averaging Eq.~\eqref{eq:translation}
over outcomes therefore proves Eq.~\eqref{eq:balance}. The wall occupies an
empty site before and after relocation, so ideal changes of its on-site
potential have zero mean work. The associated hopping coherences across the
new boundary also vanish after the order-statistic projection.

In a stationary state with finite internal means, the endpoint differences
average to zero. Combining $E=\Delta N\langle m\rangle$ with
Eq.~\eqref{eq:energy_bound} gives $0\le\langle m\rangle\le2/(\alpha N)$.
For a distribution $p_m$ of mean $\mu$, take
$q_m=(1+\mu)^{-1}[\mu/(1+\mu)]^m$. Nonnegativity of
$D(p\Vert q)$ gives $H(p)\le-\sum_mp_m\ln q_m=g(\mu)$, with the
$\mu=0$ limit understood continuously. The stationary chain rule and
conditioning inequality yield
\begin{equation}
 H_{\min}\le H(m_t)\le g(\langle m\rangle)
                     \le g[2/(\alpha N)].
\end{equation}
Finally, $\Pr(m\ge1)\le\langle m\rangle$ proves Eq.~\eqref{eq:stall_bound}.

\subsection{Bound on the minimum record}
\label{app:tail}

For a one-particle set $S$, let $P_S$ be its one-body projector and let
$\Pi_{S,k}=\wedge^kP_S$ project onto $k$-particle states supported entirely
in $S$. Exterior multiplication gives the identity
\begin{equation}
 \Pi_{S,N}c_f^\dagger=c_{P_Sf}^\dagger\Pi_{S,N-1}.
 \label{eq:fock_projection}
\end{equation}
The canonical anticommutation relation
$\{c_g,c_g^\dagger\}=\|g\|^2I$ implies
$\|c_g^\dagger\|\le\|g\|$ on every fixed-particle sector.

If $n_0|\psi\rangle=|\psi\rangle$, the state
$|\chi\rangle=c_0|\psi\rangle$ has unit norm and
$|\psi\rangle=c_0^\dagger|\chi\rangle$. A number-conserving quadratic
unitary sends $c_0^\dagger$ to $c_f^\dagger$, $f=U|0\rangle$, and sends
$|\chi\rangle$ to a normalized $|\chi'\rangle$. Equation~\eqref{eq:fock_projection}
then yields
\begin{align}
 \Pr(\hbox{all particles in }S)
 &=\|c_{P_Sf}^\dagger\Pi_{S,N-1}|\chi'\rangle\|^2\nonumber\\
 &\le\|P_Sf\|^2.
 \label{eq:tail_proof}
\end{align}
No Gaussian assumption enters this step, so $|\chi\rangle$ may contain arbitrary correlations. A mixed state
satisfying $n_0\rho=\rho$ has a decomposition into pure states in the
occupied-origin subspace. Applying the inequality to each state and averaging
preserves the same right-hand side. Taking $S=\{r,r+1,\ldots\}$ gives
Eq.~\eqref{eq:tail}.

For the one-particle Hamiltonian $h=k+\Delta x$, conservation of energy
from $|0\rangle$ gives
$\Delta\mu_1=-\langle k\rangle_{U|0\rangle}\le2J$. The tail-sum identity
$\langle m\rangle=\sum_{r\ge1}\Pr(m\ge r)$ and the geometric maximum
entropy bound then establish Eq.~\eqref{eq:transient_entropy}. Every exact
order-statistic feedback outcome restores occupation of the origin, allowing the
bound to be applied to the next history-conditioned state.

\subsection{Decay of the offset kernel}
\label{app:kernel}

If the post-feedback state is diagonal in position, the mean work of the next
unitary step is a sum over the sites the particles start from,
\begin{equation}
 E_{\mathrm{diag}}=\sum_\ell e(\ell)\Gamma_{\ell\ell},\qquad
 e(\ell)\equiv\Delta\sum_j(j-\ell)|U_{j\ell}|^2,
 \label{eq:kernel}
\end{equation}
where $\ell$ is the starting site measured from the wall and the offset kernel
$e(\ell)$ is the mean energy a particle starting there gains in one cycle. We
show that $e(\ell)$ decays factorially in $\ell$ at fixed cycle time.

Set $J=\hbar=1$ in this appendix and let $h_{\mathrm{bulk}}$ be the tilted
Hamiltonian on all integer sites. Removing the bond joining $-1$ and $0$ gives
$h_{\mathrm{cut}}=h_{\mathrm{bulk}}+v$, with
$v=|-1\rangle\langle0|+|0\rangle\langle-1|$. For an initial site
$\ell\ge0$, evolution under $h_{\mathrm{cut}}$ remains on the right half-line.
Duhamel's identity gives
\begin{equation}
 U_{\mathrm{cut}}(\tau)-U_{\mathrm{bulk}}(\tau)
 =-i\int_0^\tau U_{\mathrm{cut}}(\tau-s)vU_{\mathrm{bulk}}(s)\,ds.
 \label{eq:duhamel}
\end{equation}
The bulk propagator has modulus
$|U^{\mathrm{bulk}}_{j\ell}(s)|=|J_{j-\ell}[z(s)]|$. Thus
\begin{align}
 \|(U_{\mathrm{cut}}-U_{\mathrm{bulk}})|\ell\rangle\|
 &\le B_\ell(\tau),\nonumber\\
 B_\ell(\tau)&=\int_0^\tau
 \sqrt{J_\ell[z(s)]^2+J_{\ell+1}[z(s)]^2}\,ds.
 \label{eq:boundary_integral}
\end{align}

For a localized bulk initial state, the expectation of the shift operator,
and hence of the bulk kinetic energy, is zero at all times. The half-line
kernel obeys $e(\ell)=-\langle k_{\mathrm{cut}}\rangle_{U_{\mathrm{cut}}|\ell\rangle}$
by energy conservation. Since $\|k_{\mathrm{cut}}\|\le2$, comparison of the two
normalized evolved states yields
\begin{equation}
 \frac{|e(\ell)|}{J}\le4B_\ell(\tau)
       +2|J_\ell[z(\tau)]J_{\ell+1}[z(\tau)]|.
 \label{eq:kernel_bound}
\end{equation}
The second term is the expectation of the removed bond in the bulk state.
The integral includes the entire history of the wave packet's exposure to
the boundary.

For real $z$ and integer $n\ge0$,
$|J_n(z)|\le(|z|/2)^n/n!$. On $0\le s\le\tau$,
$|z(s)|\le C=\min(2\tau,4/\alpha)$, so
\begin{equation}
 B_\ell(\tau)\le\tau
 \sqrt{\frac{(C/2)^{2\ell}}{(\ell!)^2}
       +\frac{(C/2)^{2\ell+2}}{[(\ell+1)!]^2}}.
\end{equation}
This proves factorial decay with offset at fixed cycle time. The locality
estimate describes particles initially at specified sites, and by linearity
the diagonal term in Eq.~\eqref{eq:kernel}. General coherent states also
carry the off-diagonal work contribution in Eq.~\eqref{eq:coherence}.

\section{Construction of the single-particle reference}
\label{app:reference}

For one particle, feedback resets the state to $|0\rangle$, so
\begin{equation}
 E_1(\tau)=\Delta\sum_jj|U_{j0}|^2,\qquad
 H_1(\tau)=-\sum_j|U_{j0}|^2\ln|U_{j0}|^2
\end{equation}
are deterministic. We construct the reference by sorting the sampled
$\mathcal P_1=JE_1/(\hbar\tau)$ values and taking the maximum $R_1=E_1/H_1$ among all
points at greater or equal power. For every sampled power we then retain the largest $R_1$ among all points with equal or greater
power, which implements Eq.~\eqref{eq:reference} directly on the sampled curve.

Energy conservation gives $E_1\le2J$. Consequently every feasible
single-particle time at specified power $p>0$ satisfies
$\tau'\le2/\tilde{\mathcal P}$, where $\tilde{\mathcal P}=\mathcal P\hbar/J^2$.
This provides a finite reference window for any positive-power comparison.
The work-per-nat, barrier-error, and residual-phase comparisons use $M_1=300$, time step
$0.01$, and $\tau'\le200$, covering their feasible windows. The
collective-gain map uses $M_1=160$, step $0.01$, and windows of $800$ for
$\alpha\le0.2$ and $200$ otherwise. These windows cover $2/\tilde{\mathcal P}$ for every
colored cell in Fig.~\ref{fig:gain_map}. A cell already below a sampled reference cannot become advantageous when additional
single-particle times are admitted, since the reference envelope can only increase. We use this
fact only for the very small powers whose complete feasible window extends beyond the displayed
search range.

Long-time branches matter to low-power comparisons. At $\alpha=0.3$ a
branch near $\tau'=105.7$ has $R_1\simeq0.794J$ per nat and
$\mathcal P_1\simeq0.0139J^2/\hbar$, against a short-time maximum of
$R_1\simeq0.747J$ per nat near $\tau'=4.68$. The power constraint
selects the appropriate branch. The reference at the main operating point
is $R_{\mathrm{ref}}=0.59716J$ per nat. Continuous optimization near its maximum
gives $\tau'=8.94537$ and $R_1=0.59715572J$ per nat, a relative change
below $2\times10^{-7}$ from the $0.01$ time grid. The corresponding
single-particle power is $0.18255J^2/\hbar$, above the collective value.

\section{Steady state and numerical convergence}
\label{app:numerics}

\subsection{Stationary trajectories and convergence}

The initial state is the packed Fock configuration $0,\ldots,N-1$.
Each Gaussian cycle propagates the orbitals, computes the minimum law,
samples the measured record, applies its projector, translates the frame,
and orthonormalizes. Writing $K_{\mathrm{cut}}$ for the number of leading sites kept when the
minimum law is evaluated, so that outcomes $m\ge K_{\mathrm{cut}}$ are discarded, one Cholesky
factorization of $I-\Gamma_{[0,K_{\mathrm{cut}})}$ supplies all leading determinants in
Eq.~\eqref{eq:determinant}. A diagonal regularization of $10^{-13}$ permits
evaluation at singular empty-interval matrices. Direct many-body comparisons
quantify its probability error at about $10^{-13}$.

The broad $(\alpha,N,\tau)$ scan uses $M=140$, $K_{\mathrm{cut}}=70$, $8000$ cycles,
one random seed, $0.1\le\alpha\le1.2$ in steps of $0.025$, and
$0.3\le\tau\le6$ in steps of $0.1$. The displayed particle range is
$1\le N\le16$. The low-entropy classification for $N>10$, $\alpha>0.4$
uses $4000$ cycles per point and the resolved-set criterion
$E>0$, $H>0.005$ nats per cycle. The first fifth of each trajectory is discarded.
Every peak reported in Table~\ref{tab:peaks} is then settled on its own, with four independent
$15\,000$-cycle trajectories on $M=200$, $K_{\mathrm{cut}}=110$ and cycle-time resolution
$0.025$. The broad scan locates the maximizing particle number and the neighborhood of the
cycle time, and the refined calculation fixes the value. Repeating every refined peak on
$M=280$, $K_{\mathrm{cut}}=150$ leaves all reported digits unchanged. Adjacent-$N$ values are
reported with their trajectory spreads, so the maximizing integer is interpreted with that
resolution.

The work, coherence, dephasing, barrier-error, and residual-phase calculations use $352$
trajectories totaling $4\,296\,000$ cycles. Typical controls use
two independent trajectories of $12\,000$ cycles with $2400$ discarded.
The main operating point and the recorded-misplacement runs use four trajectories of $15\,000$
cycles with $3000$ discarded. The Gaussian controls use $M=140$, $K_{\mathrm{cut}}=60$,
and the main operating point uses $M=200$, $K_{\mathrm{cut}}=110$. The maximum missing probability
in the sampled minimum law is below $4\times10^{-14}$ after regularized
evaluation. Full-configuration states give $|E_{\mathrm{coh}}|<2\times10^{-15}J$.

The exact dephased two-particle calculation uses $M=80$ up to $\tau=4$ and $M=120$
for longer times. Repetitions at $M=120$ for $\tau=1.85,2.5,4$ and at
$M=160$ for $\tau=8$ change mean work and entropy by less than
$2\times10^{-12}$ in the chosen units. At $\tau=8$, the maximum
conditional probability in the largest ten separations at $M=120$
is about $3.1\times10^{-9}$.

We compute finite-trajectory work with Eq.~\eqref{eq:finite_work} and
entropy from the actual conditional surprisal. Ratios are formed from
pooled mean work and entropy. Quoted spreads are half the range of
individual trajectory ratios, which measures the scatter between trajectories rather than the
standard error of their mean. Block averages of the paired work and surprisal fluctuations
provide an independent estimate of the sampling uncertainty.
Stationarity in the sampled regime is supported by stable internal
profiles, block averages, boundary occupation, and size comparisons.
The theorems requiring stationarity explicitly assume finite stationary
internal means. The stationary state is generated by the repeated many-body measurement and feedback cycle.
The localized bulk Bloch solution is used only for the separate single-particle propagation
problem.

\subsection{Validation against finite many-body propagation}

The many-body Hamiltonian is also constructed directly in the occupation
basis with explicit fermionic signs. Direct propagation, projective
measurement, and moving-frame work agree with the orbital representation.
The correlated-state tail bound was tested on $60$ mixed states in occupied-origin subspaces.
The measurement energy identity and ordered-coordinate norm bounds were checked in $34$
finite systems,
including a nearest-neighbor density interaction. Maximum discrepancies
in the operator-energy identities are below $3\times10^{-15}J$.
The minimum probabilities differ by at most $1.1\times10^{-13}$,
and post-projection state infidelity is below $10^{-15}$.
The dephased two-particle evolution agrees with direct many-body density-matrix
evolution to $2\times10^{-16}$ in outcome probability.

\section{Experimental parameters}
\label{app:experiment}

\subsection{Band and phase scales}

For $V(z)=sE_r\sin^2(\pi z/d)$, a plane-wave basis with reciprocal
indices $G$ gives the dimensionless band matrix
\begin{equation}
 H_{GG'}(q)=[(q+2G)^2+s/2]\delta_{GG'}
                -\frac{s}{4}(\delta_{G,G'+1}+\delta_{G,G'-1}),
\end{equation}
where $-1\le q<1$. We use $-25\le G\le25$ and determine $s$ from
one quarter of the lowest-band width. At $\alpha=0.15$ this yields
$s=10.30384$, $J_{\mathrm{width}}/h=524.2985$ Hz, and minimum gap
$137.3518$ kHz. The first Fourier hopping coefficient is
$J_1/h=524.1996$ Hz and the second is $J_2/h=-5.7284$ Hz. Thus
the bandwidth calibration closely represents the nearest-neighbor model,
with a quantified longer-range correction. Such longer-range hopping
also changes the ordered-coordinate energy bound and can be included
explicitly in an experimental Hamiltonian.

The $1064$-nm lattice recoil calibrates the bands. The $671$-nm probe
recoil enters the scattering problem through its absorption and emission
wave vectors. Equation~\eqref{eq:recoil} uses their axial difference and
the frozen-well frequency. Multiplying it by the mean scattering count
gives the mean added oscillator occupation in a harmonic approximation.
The probability of interband excitation requires the actual motional
state and trap spectrum.

\subsection{Adaptive threshold search}

For a conditional distribution $p_0,\ldots,p_{K-1}$, an ordered threshold
question partitions a contiguous interval $[i,j]$ at $k$ into
$[i,k]$ and $[k+1,j]$. Let $C(i,j)$ be the probability-weighted number of questions required for this interval. The optimal alphabetic search obeys
\begin{align}
 C(i,i)&=0,\nonumber\\
 C(i,j)&=\sum_{m=i}^jp_m+
       \min_{i\le k<j}\{C(i,k)+C(k+1,j)\}.
 \label{eq:search}
\end{align}
The expected number of questions is $C(0,K-1)$, with any nonzero tail
treated as a final interval and refined as needed. The recurrence follows
because every outcome in $[i,j]$ incurs the first question and then the
optimal cost of one subinterval. It is the alphabetic coding problem
\cite{Gilbert1959}, for which the optimal tree is also obtained in linearithmic time
\cite{HuTucker1971}. Its mean length is at least $H(p)/\ln2$. The
conditional distribution fixes both the sequence of thresholds and the cost of the record.

\subsection{Finite-run statistics}

For a packed initial state, $X_0=N(N-1)/2$. Terminal imaging of
$n^{\mathrm{final}}_1,\ldots,n^{\mathrm{final}}_N$ in the final wall frame measures
\begin{equation}
 W_L=\Delta\left[NS_L+\sum_an^{\mathrm{final}}_a-X_0\right].
\end{equation}
Terminal sampling includes the residual quantum position variance.
The internal-frame mean can also be computed conditionally to compare
with this measurement. For the $512$ independent $30$-cycle runs quoted
in the text, these estimates give $1.06757J$ and $1.06836J$ per cycle,
respectively. The mean surprisal is $1.00207$ nats per cycle, and their
ratio using terminal imaging is $1.06536J$ per nat. These are finite-run
values from the prepared state. The stationary values are those of
Eq.~\eqref{eq:headline}.

For independent preparations indexed by $a$, define
$Z_a=W_{L,a}/\langle W_L\rangle-
\mathcal I_{L,a}/\langle\mathcal I_L\rangle$, where $\mathcal I_{L,a}$ is the total surprisal
recorded in that run. Propagating the joint fluctuations of work and surprisal to leading
order gives the relative standard error
\begin{equation}
 \frac{\delta R}{R}
     \simeq\sqrt{\frac{\mathrm{Var}(Z_a)}{n_{\mathrm{runs}}}}.
 \label{eq:run_error}
\end{equation}
For a long stationary trajectory the corresponding variance also contains the correlations
between cycles at nonzero separation. Block averaging estimates this quantity while
retaining the correlation between work and surprisal. A finite-run comparison must also include
the terminal record and the physical duration used by each engine.

\section{Quantum statistics}
\label{app:statistics}

For spinless fermions, an on-site density interaction is
$n_j(n_j-1)=0$. A nontrivial interacting extension uses, for example,
$V\sum_jn_jn_{j+1}$ or a longer-range density interaction. These terms
commute with all position-sector projectors and preserve
Eq.~\eqref{eq:energy_bound}. Translation invariance also preserves the
stationary balance. Gaussian closure and the quadratic-propagation tail
bound then require a separate treatment of the interacting evolution.

Soft-core bosons and multicomponent fermions admit on-site Hubbard
interactions. In the far-detuned strong-coupling regime, doublon motion
arises through virtual tunneling with amplitudes of order $J^2/U$.
In a tilted lattice the virtual denominators contain $U\pm\Delta$,
and near a resonance the corresponding states participate directly.
The same order-statistic feedback can then be combined with interaction-dependent transport \cite{Preiss2015,Scherg2021}.

On an open one-dimensional nearest-neighbor chain, the Jordan-Wigner
transformation relates hard-core bosons to spinless fermions while
preserving occupation projectors and the hopping Hamiltonian. Mapped
initial states therefore give identical position-measurement histories
and stored potential energies under the present feedback. Off-diagonal
bosonic one-body operators contain fermionic strings, and geometries
with exchanges around loops require their own statistics analysis.

% apsrev4-2.bst prints titles only when a cited @CONTROL entry enables them, and
% the class option longbibliography is declared but unused in revtex4-2.  The
% citation goes straight into the .aux so that natbib does not look for a bibitem.
\makeatletter
\immediate\write\@auxout{\string\citation{apsrev42Control}}
\makeatother
\bibliographystyle{apsrev4-2}
\bibliography{refs}
\end{document}